\documentclass[useAMS,usenatbib]{mn2e}
\usepackage{graphicx}
\usepackage{txfonts}
   \title[CCD $uvby\beta$ photometry of NGC 663]{CCD 
    \textit{uvby}$\bmath{\beta}$ photometry of the young open cluster 
    NGC 663}

   \author[J. Fabregat and G. Capilla]{J. Fabregat\thanks{E-mail: 
juan.fabregat@uv.es} and G. Capilla 
\\
   Observatorio Astron\'omico,
              Universidad de Valencia, 46100 Burjassot,
        Spain} 
\begin{document}

\date{Accepted date. Received date; in original form date}

\pagerange{\pageref{firstpage}--\pageref{lastpage}} \pubyear{2004}

       \maketitle

\label{firstpage}

\begin{abstract}
   NGC 663 is a young, moderately rich open cluster, known to contain 
   one of the largest fractions of Be stars among all galactic clusters. 
   In this work we present CCD $uvby\beta$ photometry for stars in its 
   central area. We have used these data to obtain the main cluster 
   physical parameters. We find that the reddening is highly variable, 
   with values ranging from $E(b-y) = 0.639\pm0.032$ in the central part 
   to $E(b-y) = 0.555\pm0.038$ in the south-east. The distance modulus is 
   found to be $11.6\pm0.1$ mag. (2.1 Kpc), and the age $\log t = 
   7.4\pm0.1$ years ($25^{+7}_{-5}$ Myr). The age obtained is consistent 
   with the interpretation of the Be phenomenon as an evolutionary effect. 
\end{abstract}


%

    \section{Introduction}

The precise determination of the main physical parameters of galactic 
open clusters plays a central role in the study of the stellar structure
and evolution. With accurate photometric data, and once the external
variables such as reddening are corrected for, the cluster distances, ages
and chemical abundances can be inferred from the study of the photometric
colour-magnitude and colour-colour diagrams. 

This work is part of a programme aimed to produce
accurate and homogeneous dating for a sample of young
galactic clusters by means of CCD Str\"omgren $uvby\beta$ photometry. 
The use of isochrone fitting in the $V_0 - c_0$ plane of the $uvby$
photometric system is an adequate tool to obtain accurate cluster ages. 
The range of variation of the $c_0$ index along the B-type sequence
amounts to more than one mag., being significantly larger than most 
commonly used photometric colours. Moreover, the $c_0$ index is less 
affected by reddening, and allows an efficient segregation of 
emission-line stars. 

NGC 663 is a moderately rich open cluster in Cassiopeia, located 
in the Perseus galactic arm. It is thought to be the core of the 
Cas OB8 association. A remarkable characteristic is that it 
contains one of the largest fractions of Be stars among all 
galactic clusters. More than 30 Be stars have been detected so 
far among its B star sequence, and at least 30 per cent of all early B 
cluster members are known to be Be stars \citep*{sand79,sand90,pig01a}.

The accurate age determination of clusters with a high Be star 
content is a key element of the on-going discussion about the 
evolutionary status 
of Be stars. \citet{fabregat00} proposed that the Be phenomenon is an 
evolutionary effect which appears at the end of the B stars main sequence 
lifetime. The main argument in favour of this hypothesis is their 
preliminary finding, based on an inhomogeneous set of age determinations 
from the literature, that clusters with high Be star content have 
ages in the interval 13--25 Myr, while no -- or very few -- Be stars 
are found in clusters younger than 10 Myr. Hence, the accurate 
determination of the age of NGC 663 is an important issue in this 
discussion.

NGC 663 has been the subject of many photometric studies in the 
past. Photoelectric $UBV$ photometry was obtained by  
\citet{hoag61}, \citet{mccuskey} and \citet{bergh}. Str\"omgren $uvby$ and 
Crawford--Barnes 
H$\beta$ photometry was presented by \citet{tapia} and \citet{fabregat96}. 
The former 
also obtained infrared $JHK$ photometry. Recent studies based on CCD 
photometry in the Johnson $UBV$ and Cousins $RI$ systems include those of 
\citet{phelps} and \citet*{pig01a,pig01b}. Photometric H$\alpha$ surveys 
aimed to detect new Be stars have been conducted by \citet*{capilla00} 
and \citet{pig01a}.

The cluster physical parameters determined in the above studies show 
an important dispersion. A reason for this could be the strongly 
variable reddening across the cluster area (e.g. \citealt{bergh}) which 
makes the determination of the intrinsic magnitudes and colours more 
difficult. The published distance moduli range from 11.4 mag. (1.9 Kpc, 
\citealt{mccuskey}) to 12.25 mag. (2.8 Kpc, \citealt{phelps}). The 
determined ages range from 9 Myr. \citep{tapia} to 20--25 Myr. 
\citep{pig01a}.

In this work we present CCD $uvby\beta$ photometry for stars in the 
central area of NGC 663 and a precise determination of the cluster main 
physical parameters. A detailed study of the cluster Be star population 
will be presented in a forthcoming paper.


\section{The data}

CCD photometry of NGC 663 was obtained on the nights 20 to 22 November 
1998 at the 1.52-m. telescope of the Observatorio Astron\'omico 
Nacional, located at the Calar Alto Observatory (Almer\'{\i}a, Spain). 
The chip employed was the Tektronics TK 1024 AB, with a size of 
1024\,x\,1024 pixels. The 0\farcs 4 unbinned pixels provide a field size 
of 6\farcm 9\,x\,6\farcm 9. Two slightly overlapping fields near the 
cluster centre were observed to cover the cluster central region.

Observations were done through the four Str\"omgren $uvby$ and
Crawford narrow and wide H$\beta$ filters, every field being
sequentially measured through the six filters. Three different
exposure times were used with each filter, in order to ensure a wide
range of stellar magnitudes. Exposure times in each filter were
selected so that a B type star produces approximately equal count rates
through all filters. Employed exposure times are presented in Table 
\ref{t1}. The list of all observations is presented in Table \ref{t2}.

\begin{table}
\centering
\caption{Exposure times with the different filters used, in seconds.}
\label{t1}
\begin{tabular}{lrrr}
\hline
Filter & short & medium & large \\
\hline
$y$ & 10 & 50 & 200  \\
$b$ & 12 & 60 & 240 \\
$v$ & 35 & 175 & 700\\
$u$ & 120 & 600 & 2$\times$1200 \\
H$\beta_{w}$ & 12 & 60 & 240 \\
H$\beta_{n}$ & 30 & 150 & 600 \\
\hline
\end{tabular}
\end{table}

\begin{table}
\centering
\caption{List of observations}
\label{t2}
\begin{tabular}{@{}cccl}
\hline
JD & date & airmass & exposure \\
\hline
2\,451\,138 & 20-11-98 & 1.26 & short\\
2\,451\,138 & 20-11-98 & 1.40 & long\\
2\,451\,139 & 21-11-98 & 1.28 & short\\
2\,451\,139 & 21-11-98 & 1.46 & long\\
2\,451\,140 & 22-11-98 & 1.15 & short\\
2\,451\,140 & 22-11-98 & 1.17 & long\\
2\,451\,140 & 22-11-98 & 1.45 & medium\\
2\,451\,140 & 22-11-98 & 1.56 & medium\\
\hline
\end{tabular}
\end{table}


During the same observing run we also obtained photometry of the  
clusters $h$ \& $\chi$ Persei, which has been published elsewhere 
(\citealt{capilla}, hereafter referred to as CF02). A 
detailed description of the image processing procedures, atmospheric 
extinction calculation and standard $uvby\beta$ transformations is given 
in CF02. The accuracy of the standard photometry is 0.023, 0.014, 
0.017, 0.017 and 0.021 mag. in $V$, $(b-y)$, $m_1$, $c_1$ and $\beta$ 
respectively. In CF02 it is also shown that our photometry is well 
tied to the standard system and free of systematic effects. 

Astronomical coordinates for all observed stars were derived from the 
instrumental pixel coordinates by using 26 stars from the Tycho 2 
Catalogue \citep{hog} included in the observed fields. 
Transformation equations were
computed by means of the {\em Starlink} program ASTROM \citep{wallace}. 
The final astrometric accuracy, measured as the RMS of
the mean catalogue minus transformed values for the stars used in the
transformation, is better than  0\farcs 1 both in Right Ascension and 
Declination.

Equatorial coordinates and mean photometric magnitudes, colours and
indices for stars in the central region of NGC 663 are presented in
Table \ref{t4}. We have adopted the cluster star numbering system from the 
WEBDA database\footnote{http://obswww.unige.ch/webda/} \citep{merm}. 
For cluster numbers lower than 635, WEBDA numbers are 
coincident with \citet{wall} numbers, which we will refer to 
as 'W' hereafter.  

\begin{table*}
\centering
\caption{Coordinates and photometric data for stars in the NGC 663 area. N 
is the number of $uvby$ and $H\beta$ individual measurements. Stars 
considered as cluster members are denoted with 'm' in the last column. 
Only a few entries are displayed. The complete table is available in 
electronic form.}
\label{t4}
\begin{tabular}{@{}rccrcrcccccccccc}
\hline
\noalign{\smallskip}
Star & $\alpha$(J2000) & $\delta$(J2000)& $V\hspace{2mm}$ & $b-y$ &
$m_1\hspace{1mm}$ & $c_1$ &
$\sigma_V$ & $\sigma_{b-y}$ & $\sigma_{m_1}$ & $\sigma_{c_1}$ &N& $\beta$ 
&
$\sigma_{\beta}$ & N & M \\
\hline
1 & 01 46 33.806 & +61 13 34.04 & 14.653 & 0.545 & --0.144 & 0.819 & 0.016
& 0.017 & 0.045 & 0.065 & 5 & 2.767 & 0.060 & 3 & m \\
2 & 01 46 35.604 & +61 13 39.08 & 12.258 & 0.572 & --0.196 & 0.230 & 0.010
& 0.007 & 0.020 & 0.021 & 6 & 2.473 & 0.017 & 3 & m \\
3 & 01 46 39.876 & +61 14 02.51 & 13.029 & 0.508 & --0.124 & 0.538 & 0.003
& 0.009 & 0.016 & 0.022 & 6 & 2.595 & 0.042 & 3 & m \\
4 & 01 46 38.997 & +61 14 06.06 & 11.025 & 0.490 & --0.166 & 0.131 & 0.002
& 0.016 & 0.036 & 0.034 & 4 & 2.558 & 0.026 & 3 & m \\
5 & 01 46 37.352 & +61 14 17.48 & 14.288 & 0.561 & --0.150 & 0.663 & 0.015
& 0.016 & 0.032 & 0.036 & 6 & 2.708 & 0.009 & 3 & m \\
\hline
\end{tabular}
\end{table*}


\begin{figure}
\resizebox{\hsize}{!}{\includegraphics{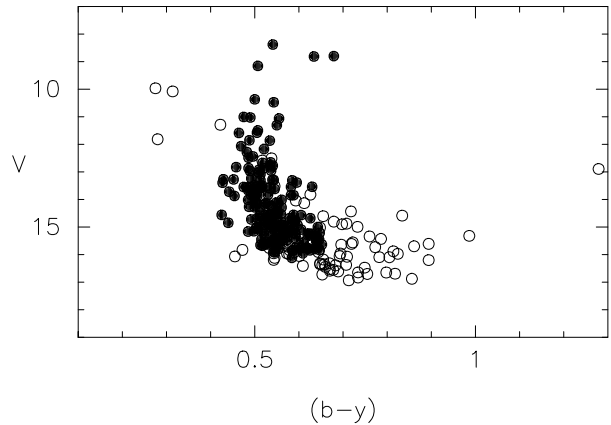}}
\caption{Colour-magnitude diagram for NGC 663. Filled
and open circles denote stars considered as cluster members and nonmembers 
respectively} 
\label{f1}
\end{figure}

\begin{figure}
\resizebox{\hsize}{!}{\includegraphics{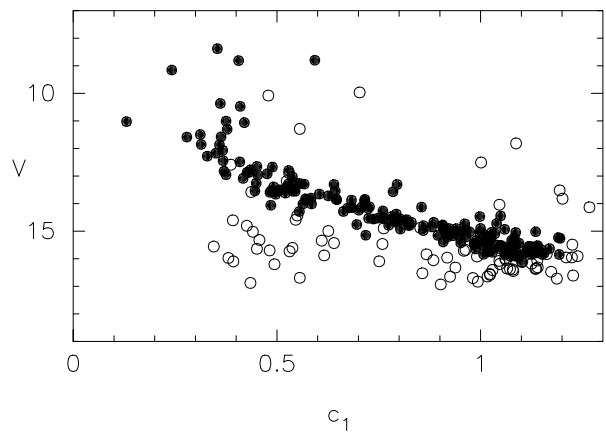}}
\caption{$V-c_1$ photometric diagrams. Symbols as in Fig.~\ref{f1}}
\label{f2}
\end{figure}

\section{Reddening, intrinsic colours and distance}

The colour-magnitude diagram of all observed stars is presented in 
Fig.~\ref{f1},  and the photometric $V-c_1$ diagram in Fig.~\ref{f2}. To 
obtain the intrinsic colours we first classified the stars as belonging 
to the early (earlier than A0), intermediate (A0-A3) or late (A3 
onward) groups defined by \citet{strom}. The 
classification was performed by means of the algorithm described by 
\citet{figueras}. 

Cluster membership was assigned from the position of each star in the
different $uvby\beta$ photometric diagrams. Stars considered as members
are marked in the last column of Table \ref{t4}, and are represented with
a different symbol (filled circle) in Figs.~\ref{f1} and \ref{f2}. The 
reliability of the photometric membership criteria was checked in 
CF02, where we conclude that our cluster photometric sequences  
are not significantly contaminated with the inclusion of nonmember field 
stars, and, conversely, very few, if any, actual members have been 
excluded from them.

All member stars belong to the early group, indicating that our 
magnitude limited photometry does not reach beyond the late B spectral 
types. Reddening values and intrinsic colours and indices were 
hence obtained by means of the procedure described by \citet{craw78}. We 
have used the standard $(b-y)_0 - c_0$ relation given
in Table VI of \citet*{perry87}, and the following
reddening relations:

\begin{quote}
$A_V = 4.3 E(b-y)$ \\
$E(c_1) = 0.2 E(b-y)$ \\
$E(m_1) = -0.32 E(b-y)$ 
\end{quote}

Known supergiant and Be stars have not been included in the computation of 
the interstellar reddening. B type supergiants do not follow the standard  
$(b-y)_0 - c_0$ relation, while Be stars present an additional reddening
contribution of circumstellar origin.

\begin{figure}
\resizebox{\hsize}{!}{\includegraphics{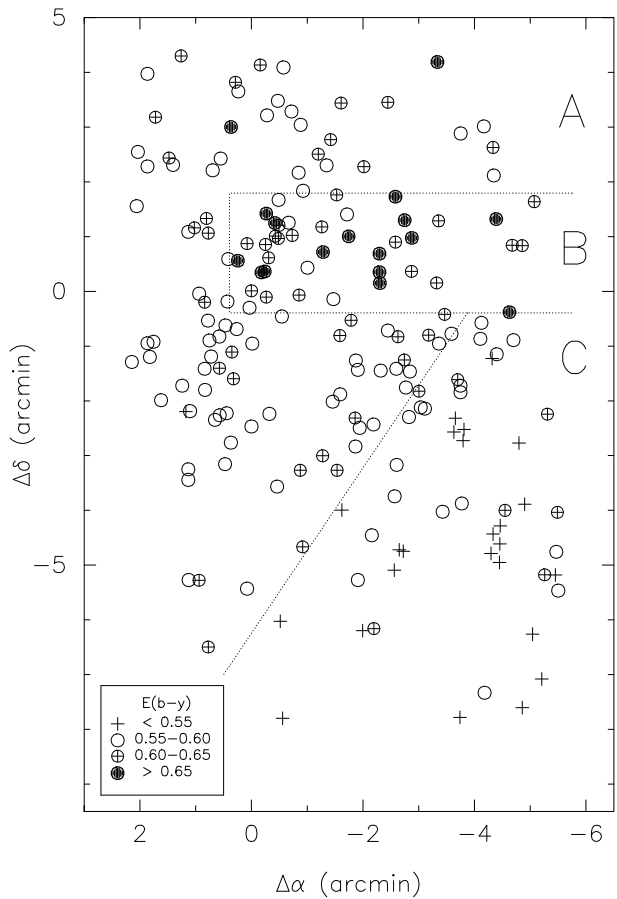}}
\caption{Reddening spatial distribution for B stars in NGC 663. 
Positions are relative to star W22 
(01$^{\rm h}$46$^{\rm m}$30\fs 24, +61\degr 12\arcmin 59\farcs 6,
J2000)}
\label{f3}
\end{figure}

We obtained a mean reddening value of $E(b-y) = 0.591 \pm
0.042$, from 189 stars. The large value of the standard deviation
indicates the presence of variable reddening across the cluster
area. In Fig.~\ref{f3} we have represented the reddening values for
individual stars as a function of their position. The cluster centre is 
the most heavily reddened region, while in the south-east part the 
extinction is significantly lower than the cluster average, as already 
noted by \citet{fabregat96}. In Fig.~\ref{f3} we
have divided the cluster nucleus area into three regions of different 
reddening. Mean reddening values in these regions are presented in 
Table \ref{t6}. Within each region, and in particular in regions B and C, the 
standard deviation of the mean reddening is still large, indicating 
that we cannot consider constant extinction within the region. 
Instead, the extinction appears to be clumpy and irregularly variable with 
high spatial frequency.

\begin{table}
\centering
\caption[]{Mean reddening values for the three regions of the NGC 663
nucleus defined in Fig.~\ref{f3}}
\label{t6}
\begin{tabular}{ccr}
\hline 
Region & $E(b-y)$ & stars  \\
\hline
A  & $0.590 \pm 0.026$ & 100 \\
B  & $0.639 \pm 0.032$ & 40 \\
C  & $0.555 \pm 0.038$ & 49 \\
\hline 
\end{tabular}
\end{table}  

\begin{figure}
\resizebox{\hsize}{!}{\includegraphics{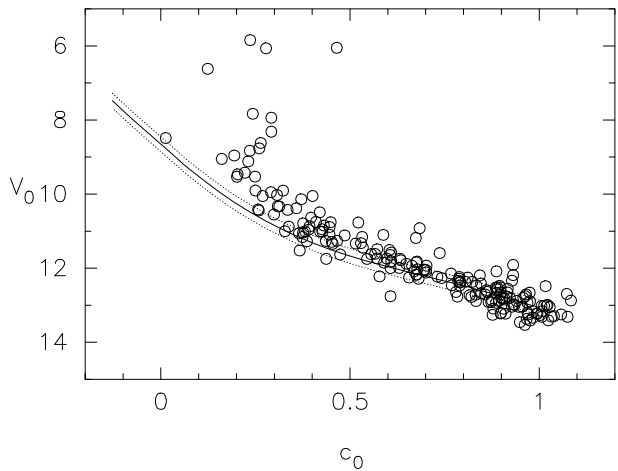}}
\caption{ZAMS fitting to the NGC 663 B star sequence. The
solid line represents the ZAMS at a distance modulus of 11.6 mag. Dotted
lines are for distance moduli of 11.4 and 11.8 respectively.}
\label{f4}
\end{figure}

In Fig.~\ref{f4} we present the intrinsic $V_0 - c_0$ diagram for B stars. 
Each star has been dereddened on the basis of its position within 
Fig.~\ref{f3}, and using the reddening values in Table \ref{t6}. 
To obtain the distance we have fitted to the $V_0 - c_0$ diagram 
the ZAMS as presented in Table VI of \citet{perry87}. We found the best 
fitting at a distance modulus of 11.6 
mag. To estimate the error of this determination we have also 
represented in Fig.~\ref{f4}, as dotted lines, 
the ZAMS shifted by distance moduli of 11.4 and 11.8 respectively. We 
find that the 11.6 value produces a distinctly better fitting than the two 
latter ones, and hence we give the value of $11.6 \pm 0.1$ mag. as the 
distance modulus of NGC 663. This value is in good agreement 
with the recent determinations based on CCD photometry.

\section{Cluster age}

Age determination has been done by isochrone fitting to
the upper main sequence. In the $uvby$ system, for early-type stars 
the $(b-y)_0$ colour and the $c_0$ index are temperature
indicators, and hence both $V_0 - (b-y)_0$ and $V_0 - c_0$ planes are
observational HR diagrams. Following the discussion in \citet{fabregat00} 
we consider the isochrone fitting to 
the $V_0 - c_0$ diagram as more precise and reliable, for the following
reasons: i./ the range of variation of the $c_0$ index along the B-type
sequence is more than ten times larger than the range of variation of
$(b-y)_0$, providing much better discrimination between isochrones of
similar ages; ii./ the $c_0$ index is less affected, by a factor of 5, by
interstellar reddening; iii./ the $V_0 - c_0$ plane allows an efficient
segregation of emission line stars.

\begin{figure}
\resizebox{\hsize}{!}{\includegraphics{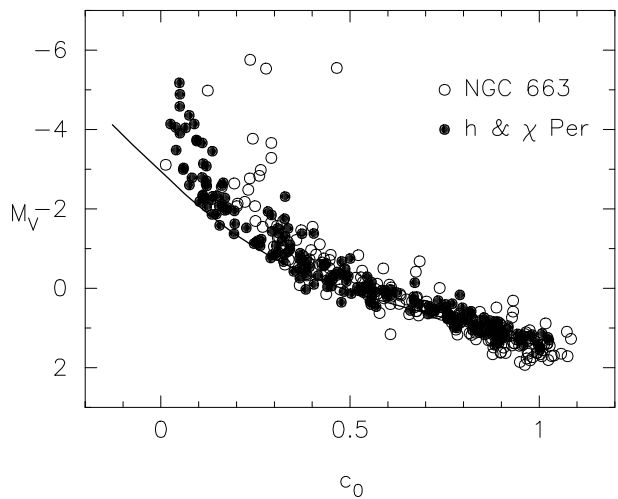}}
\caption{Comparison between the B star sequences of NGC 663 and $h$ \& 
$\chi$ Persei .}
\label{f5}
\end{figure}

As a first approach, in Fig.~\ref{f5} we compare the B star sequence of 
NGC 663 in the $V_0 - c_0$ plane with that of $h$ and $\chi$ Persei. 
Photometric data for $h$ and $\chi$ Per have been obtained from CF02. The 
turn-off of both sequences are clearly seen, and from them it is apparent 
that NGC 663 is significantly older than $h$ and 
$\chi $ Persei. Five NGC 663 stars are distinctly 
outside the cluster sequence. Star W4 lies at the 
left of the cluster turn-off, close to the ZAMS, and it is hence a cluster 
blue straggler. Its spectral type, derived from its $c_1$ index, is B1V. 
This star is also a $\beta$ Cephei pulsating variable \citep{piet,pig01b}. 
Its position in the 
photometric diagram is close to W161, another main sequence blue 
straggler already noted by \citet{fabregat96}, but 
outside the field observed in this study. 

The other four stars are in an horizontal sequence with $M_V$ between $-$6 
and $-$5 magnitudes. They are stars W40 (BD +60 343), 
W44 (BD +60 339), W54 (BD +60 333) and W86 (BD +60 331). All of them are 
supergiants. They will not be considered for isochrone fitting in the 
$M_V - c_0$ plane since the $c_0$ index defines 
different effective temperature scales for main sequence and supergiant 
stars. In CF02 there is a detailed discussion on this subject.

\begin{figure}
\resizebox{\hsize}{!}{\includegraphics{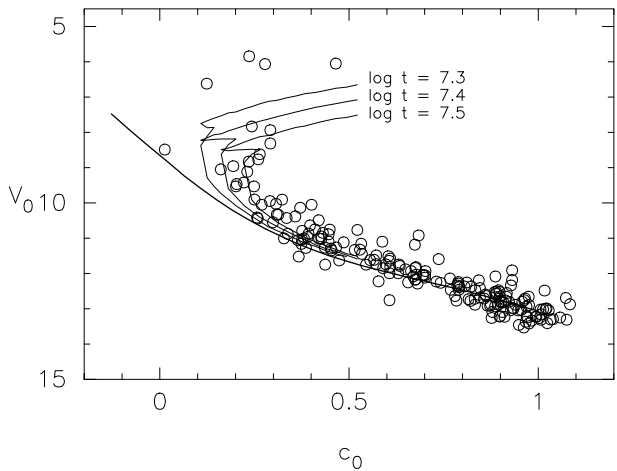}}
\caption{Isochrone fitting to the NGC 663 B star
sequence, in the $V_0 - c_0$ plane.} 
\label{f6}
\end{figure}

In Fig.~\ref{f6} we present the $V_0 - c_0$ sequence 
together with isochrones with ages of $\log t =$ 7.3, 7.4 and 7.5 
years. The isochrones have been computed with the evolutionary models of
\citet{schaller}, and transformed  to the observational
plane by means of the relations obtained  by \citet{torrejon}.   
Practically all stars at the cluster turn-off lie between the $\log t =$ 
7.4 and 7.5 isochrones. Most stars are located on the 7.5 isochrone, but 
a significant number, 5 stars, are at the left, making the 7.4 isochrone 
better for defining the lower envelope of the stellar sequence.

\begin{figure}
\resizebox{\hsize}{!}{\includegraphics{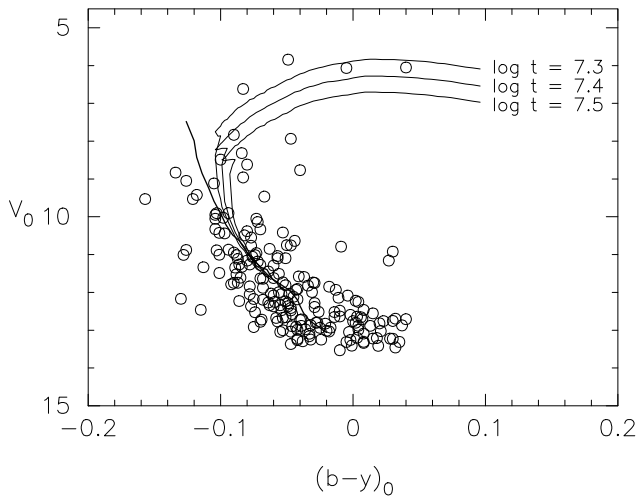}}
\caption{$V_0 - (b-y)_0$ plane for B type stars in NGC 663.}
\label{f7}
\end{figure}

In order to use the supergiant stars to further restrain the age 
determination, in Fig.~\ref{f7} we present the $V_0 - (b-y)_0$
photometric diagram. $(b-y)$ is a measure of the Paschen continuum slope,
which is correlated with effective temperature for stars of all luminosity
classes. In this plane the four supergiants are closer to the $\log t =$ 
7.4 years isochrone than to the 7.5 one. Their positions even suggest a 
still younger age. However, for the reasons discussed above, the error in 
the position of a star in the $V_0 - (b-y)$ plane is likely to be affected 
by much bigger errors, and hence we give a higher weight to the age 
determination through the $V_0 - c_0$ plane. 

From all this discussion we finally adopt an age of $\log t = 7.4\pm0.1$ 
years ($25^{+7}_{-5}$ Myr). The quoted error is higher than the one we 
estimated in our age determination of the $h$ and $\chi$ 
Persei clusters 
in CF02. The reason for this is the more variable and clumpy nature of 
the interstellar reddening in the NGC 663 field, which makes the 
reddening determination more imprecise and the sequences in the intrinsic 
photometric planes more disperse. 

\section{Discussion}

\citet{fabregat00} proposed that the Be phenomenon is an evolutionary 
effect, appearing at the end of the main sequence lifetime of a Be star. 
They based their conclusion in the analysis of the frequency of Be stars 
in clusters of different ages. They found that in clusters younger than 10 
Myr Be stars are lacking or very scarce. The maximum Be star frequency 
occurs in clusters in the 13--25 Myr age interval. The main sequence 
lifetime of a B2 star -- the subtype in which the Be star frequency reach 
its maximum -- is 26 Myr. The lack of Be stars in the youngest 
clusters indicates that a Be star cannot be a young object, but an object 
close to the end of its life in the main sequence. 

However, the above results were drawn from inhomogeneous sets of data from 
the literature. In particular, published ages are
often affected by important uncertainties. The accurate and homogeneous 
dating of young open clusters with different Be star content is hence a 
key issue to check the evolutionary hypothesis for the Be phenomenon.    

The obtained age and associated error safely exclude an age lower 
than 10 Myr, and places NGC 663 in the age interval (13--25 Myr) where the 
maximum Be star frequency occurs. A similar result was obtained in CF02 
for the clusters $h$ and $\chi$ Persei, which also have a high Be star 
content. On the other hand, in a recent paper, \citet{fu} failed to find 
any Be star in the younger cluster Trumpler 24. The results of this work, 
together with the two referred to above, are pieces of evidence supporting 
the view that the Be star phenomenon is an evolutionary effect appearing 
at the second half of the main sequence lifetime of a B star.

\section{Conclusions}
 
We have presented CCD $uvby\beta$ photometry for stars in the central
area of NGC 663. We have obtained the cluster
astrophysical parameters from the analysis of the B type star range.
The reddening is highly variable, with values ranging from $E(b-y) = 
0.639\pm0.032$ in the central part to $E(b-y) = 0.555\pm0.038$ in the 
south-east. The distance modulus is found to be 11.6$\pm$0.1 mag. 
(2.1 Kpc), and the age $\log t = 7.4\pm0.1$ years ($25^{+7}_{-5}$ Myr). 

A detailed analysis of the Be star population of NGC 663, the distribution 
of Be stars along the main sequence and their frequency as a function of 
the spectral subtype will be presented in a forthcoming paper.

\section*{Acknowledgments}
We are grateful to the Observatorio Astron\'omico Nacional for the 
allocation of observing time in the 1.5m. telescope, and for support 
during observations. 
This research has made use of the WEBDA database, developed and maintained
by J.C. Mermilliod, the SIMBAD database, operated at CDS, Strasbourg,
France, and the NASA Astrophysics Data System Abstract Service.
This work has been partially supported by the {\it Plan Nacional de
Investigaci\'on Cient\'\i fica, Desarrollo e Innovaci\'on Tecnol\'ogica
del Ministerio de Ciencia y Tecnolog\'\i a} and FEDER, through contract
AYA2000-1581-C02-01, and by {\it Generalitat Valenciana} AVCyT through 
grant GRUPOS03/170.

 \end{document}